\documentclass{aa}
\usepackage{hyperref}
\usepackage[varg]{txfonts} 
\usepackage{natbib}
\usepackage{graphicx}
\usepackage{amsmath}
\usepackage{ulem}
\usepackage{tabularx}
\usepackage[dvipsnames]{xcolor}
\newcolumntype{L}{>{\raggedright\arraybackslash}X}

\begin{document}

\title{The splashback radius  and the radial velocity profile of galaxy clusters in IllustrisTNG} 

\author{Michele Pizzardo \inst{\ref{1}}\thanks{\email{michele.pizzardo@smu.ca}}
\and Margaret J. Geller \inst{\ref{2}}
\and Scott J. Kenyon \inst{\ref{2}}
\and Ivana Damjanov \inst{\ref{1}}
}

\institute{\label{1}Department of Astronomy and Physics, Saint Mary's University, 923 Robie Street, Halifax, NS-B3H3C3, Canada
\and \label{2}Smithsonian Astrophysical Observatory, 60 Garden Street, Cambridge, MA-02138, USA
} 

\date{Received date / Accepted date}

\abstract
{
{We use 1697 clusters of galaxies from the Illustris TNG300-1 simulation (mass $M_{200c}>10^{14}$M$_\odot$ and redshift range $0.01\leq z \leq 1.04$) to explore the physics of the cluster infall region.} We use the average radial velocity profile derived from simulated galaxies, ${\rm v_{rad}}(r)$, and the average velocity dispersion of galaxies at each redshift, ${\rm \sigma_v}(r)$, to explore cluster-centric dynamical radii that characterize the cluster infall region. We revisit the turnaround radius, the limiting outer radius of the infall region, and the radius where the infall velocity has a well-defined minimum.  We also explore two new characteristic radii: (i) the point of inflection of ${\rm v_{rad}}(r)$ that lies within the velocity minimum, and (ii) the smallest radius where ${\rm \sigma_v}(r)$ = $|{\rm v_{rad}}(r)|$. These two, nearly coincident, radii mark the inner boundary of the infall region where radial infall ceases to dominate the cluster dynamics. 
{Both of these galaxy velocity based radii lie within $1\sigma$ of the observable splashback radius. The minimum in the logarithmic slope of the galaxy number density is an observable proxy for the apocentric radius of the most recently accreted galaxies, the physical splashback radius. 
The two new dynamically derived  radii relate the splashback radius to the inner boundary of the cluster infall region. }
}


\maketitle

\section{Introduction}\label{sec:introduction} 

Clusters of galaxies are massive self-gravitating  systems of galaxies comprised of a dense central region in approximate virial equilibrium surrounded by an extended region where continuing infall dominates the dynamics. The infall region extends to scales  $\lesssim 10$~Mpc \citep{Geller_1999,Rines2006CIRS,Rines2013HeCS,umetsu2014clash,Umetsu16,Umetsu20}. Across this large volume, clusters present distinctive dynamical regimes. Characteristic dynamical radii mark the transition between the virial and infall regions. These radii inform  comparisons with models of formation and evolution of cluster of galaxies \citep[e.g.,][]{press1974formation,white1978,bower1991,laceyCole93,sheth2002,zhang2008,corasaniti2011,desimone2011,achitouv2014,musso2018}.

{In standard cosmology, the turnaround radius is the outer boundary of a cluster;  it marks the radius where matter decouples from the Hubble flow. Within the turnaround, the gravitational potential of the cluster dominates the dynamics \citep{Gunn1972,silk1974,Schechter80,Bertschinger1985,villumsen86,Cupani08,Cupani11}.} Because the matter overdensity is small averaged over the volume within the turnaround radius, quasi-linear dynamics predicts the turnaround radius accurately.

Within the turnaround radius, clusters have an extended accretion region often referred to as the infall region. Here, there is a net radial infall of matter onto the cluster. {The radial velocity in this region is negative, and the well-defined minimum of the radial velocity profile is a proxy for the radius where the infall is strongest \citep{Cuesta08,deBoni2016,Valles20,pizzardo2020,Fong21,Gao23,Pizzardo23mar}. \citet{Fong21} and \citet{Gao23} recently proposed two additional characteristic radii in the infall region. They define an inner depletion radius where the mass flow rate is maximum. This radius is similar to the radius where the radial velocity has a minimum. They also define a characteristic depletion radius at slightly larger cluster-centric distance which traces the region where clustering around the halo is weakest relative to clustering around a random matter particle.}

In the dense central regions of galaxy clusters, orbital motions dominate infall. The limiting radius of this approximately virialized region, the virial radius \citep[e.g.][]{Gunn1972,Peebles80,laceyCole93}, typically defines the cluster size. Proxies for the virial radius include $R_{200c}$ and $R_{200m}$, where $R_{\Delta}$ is the cluster-centric distance that encloses a mean density $\Delta_c$ times the critical density or $\Delta_m$ times the mean background density \citep{laceyCole93}. 

A more recently suggested boundary of the inner region of a cluster is the splashback radius, $R_{\rm spl}$, {defined as} the first apocenter of orbits of recently accreted material \citep{adhikari2014,Diemer2014,More2015,Diemer20}. {The splashback radius marks a natural boundary where the internal cluster dynamics dominates over systematic infall.} 
Detailed studies use N-body simulations to explore the trajectories of individual dark matter particles as a theoretical proxy of $R_{\rm spl}$ \citep[e.g.][]{Diemer2014,Diemer2017sparta2,Mansfield17,Diemer18,Xhakaj2020}. {Usually, $R_{\rm spl}$ is outside the virialized radius of a cluster, $\sim 1.5-2 R_{200c}$.} At fixed redshift, $R_{\rm spl}$ decreases with increasing halo mass accretion rate. At fixed mass accretion rate, $R_{\rm spl}$ increases with increasing redshift. 

\citet{adhikari2014} show that at $\sim R_{\rm spl}$ particles generate caustics in phase-space density. They match this caustic to a sudden drop in the logarithmic derivative of the mass density profile of halos \citep{Diemer2014}. {This feature in the mass (or number) density profile of halos, $R_{\rm spl}^{\rho}$ ($R_{\rm spl}^{\rm n}$), is an observable proxy for $R_{\rm spl}$.} In fact, a variety of observations detect the minimum in the logarithmic slope of cluster density associated with $R_{\rm spl}$ \citep{More16,Baxter17,Chang_2018,Shin19,Zurcher19,Murata20,Adhikari21,Bianconi21,Gonzalez21}.

{So far, the physical meaning of $R_{\rm spl}$ relies solely on the dynamics of dark matter particles from pure N-body simulations. Studies based on hydrodynamical simulations investigate $R_{\rm spl}$ through its proxies, $R_{\rm spl}^{\rho}$ or $R_{\rm spl}^{\rm n}$ \citep{Baxter21,Deason21,oneil21,Dacunha22,Oneil22}. Here we use the IllustrisTNG hydrodynamical simulations \citep{Pillepich18,Springel18,Nelson19} to interpret $R_{\rm spl}$ based on the dynamics of simulated galaxies rather than dark matter particles. Connecting $R_{\rm spl}$ directly to galaxy dynamics provides a way to unveil how galaxies as well as matter particles show clear signatures of the splashback physics.  }

The mean radial velocity profile of clusters in IllustrisTNG provides additional dimensions to the view of $R_{\rm spl}$. The $R_{\rm spl}$ based on the matter density profile lies just within the radius where the radial velocity profile has a clear minimum. The radial velocity profile enables determination of two further characteristic radii: (i) the point of inflection inside the velocity minimum, and (2) the smallest radius where the local velocity dispersion exceeds the infall. 
{From the turnaround radius inward, these two radii describe the radius where infall no longer dominates the dynamics. These two radii are essentially equal to the observable proxy of $R_{\rm spl}$. This agreement extends the meaning of $R_{\rm spl}$ as the dynamical outer boundary of the region where orbital motions, as opposed to infall, dominate galaxy cluster dynamics.}

{Currently, these two galaxy radial velocity based radii are not directly observable. However, \citet{Odekon22} show that data  including independent distance measurements for galaxies around clusters and filaments allow extraction of the galaxy velocity field  in the infall region.}

Sect. \ref{subsec:cats} describes the IllustrisTNG simulations and the resulting cluster samples. Sect. \ref{subsec:profiles}  discusses the mass and velocity profiles. Sect. \ref{sec:radiigal} outlines the  definition of cluster dynamical radii based on the radial velocity profile. 
Sects. \ref{subsec:turn} and \ref{subsec:rvmin} summarize the main properties of the turnaround radius and the  radial velocity minimum, respectively. In Sect. \ref{subsec:rsplash} we identify radii based on the radial velocity profile that closely approximate the observable proxy of $R_{\rm spl}$ from the galaxy number density profile. We discuss the impact of the mass distribution on the dynamical radii in Sect. \ref{subsec:massfz}, compare the galaxy and total matter velocities in Sect. \ref{subsec:bias}, and conclude in Sect. \ref{sec:conclusion}. Table \ref{table:list_radii} defines the symbols used throughout the paper.

\begin{table}[htbp]
\begin{center}
\caption{\label{table:list_radii} Symbol definitions.}
\begin{tabularx}{\columnwidth}{l|L}
\hline
\hline
Symbol & Description \\
\hline
$\rm n_g(r)$ & average galaxy number density profile \\
${\rm v_{rad}}(r)$ & average galaxy radial  velocity profile \\
${\rm \sigma_v}(r)$ & average galaxy velocity dispersion \\
\hline
$R_{\rm spl}$ & splashback radius \\
$R_{\rm spl}^{\rm n_g}$ & minimum of the logarithmic derivative of $\rm n_g$, {a proxy of $R_{\rm spl}$} \\
$R_{\rm infl}$ & inflection point of ${\rm v_{rad}}(r)$ \\
$R_{\rm \sigma_v}$ & smallest cluster-centric distance where ${\rm \sigma_v(r)/v_{rad}}(r)=-1$ \\
\hline
$R_{\rm v_{min}}$ & cluster-centric radius of the minimum ${\rm v_{rad}}(r)$ \\ 
$R_{\rm turn}$ & turnaround radius  \\
$R_{\rm turn}^{\rm Meik}$ & turnaround radius from Meiksin approximation \\
\hline
$R_{\rm spl}^{\rm n_g,all}$ & $R_{\rm spl}^{\rm n_g}$ for all matter components \\
$R_{\rm infl}^{\rm all}$ & $R_{\rm infl}$ for all matter components\\
$R_{\rm v_{min}}^{\rm all}$ & $R_{\rm v_{min}}$ for all matter components\\
$R_{\rm turn}^{\rm all}$ & $R_{\rm turn}$ for all matter components\\
\hline
 \end{tabularx}
 \end{center}
\end{table} 

\section{Cluster samples, velocity, and mass profiles}\label{sec:vprof}

We build a sample of galaxy clusters from the TNG300-1 run of the IllustrisTNG simulations \citet{Pillepich18,Springel18,Nelson19}. We derive mass and radial velocity profiles for these clusters. We briefly discuss the cluster sample in Sect. \ref{subsec:cats} and the profiles in Sect. \ref{subsec:profiles}.

\subsection{Simulations and catalogs}\label{subsec:cats}

The IllustrisTNG simulations \citep{Pillepich18,Springel18,Nelson19} are a set of gravo-magnetohydrodynamical simulations based on the $\Lambda$CDM model. Table \ref{table:tng_det} lists the cosmological parameters of the simulations. 
TNG300-1 is the baryonic run with the highest resolution among the runs with the largest simulated volumes. The simulation has a comoving box size of $302.6$~Mpc. TNG300-1 contains $2500^3$ dark matter particles with mass $m_{\rm DM} = 5.88 \times 10^{7}~\rm{M_{\odot}}$ and the same number of gas cells with average mass $m_b = 1.10\times 10^{7}~\rm{M_{\odot}} $. 

\begin{table}[htbp]
\begin{center}
\caption{\label{table:tng_det} Cosmological parameters for IllustrisTNG.}
\begin{tabular}{llc}
\hline
\hline
Parameter & Description & Value \\
\hline
$\Omega_{\Lambda 0}$ & cosmological constant & 0.6911 \\
$\Omega_{m0}$ & total matter density & 0.3089 \\
$\Omega_{b0}$ & baryonic matter density & 0.0486 \\
$H_0$ & Hubble constant & $67.74$~km~s$^{-1}$~Mpc$^{-1}$ \\
$\sigma_8$ & power spectrum norm. & 0.8159 \\
$n_s$ & power spectrum index & 0.9667 \\
\hline
 \end{tabular}
 \end{center}
\end{table}  

As in \citet{Pizzardo23} we  use group catalogues compiled by the IllustrisTNG Collaboration to extract all of the Friends-of-Friends (FoF) groups in TNG300-1 with $M_{200c} > 10^{14}$M$_\odot$. There are 1697 clusters in the 11 redshift bins in the range $0.01\leq z\leq 1.04$. Table \ref{table:3dinfo} summarizes the main properties of the 11 subsamples, including redshift, number of clusters, median, interquartile range, and the minimum and maximum  of the mass $M_{200c}$ in each bin. 

{The mass and redshift sampling we adopt is consistent with the selection in recent studies that use hydrodynamical simulations to explore the outer region of galaxy clusters \citep[e.g][]{Baxter21,Deason21,oneil21,Dacunha22,Oneil22}. This sampling mimics characteristics of observational approaches that focus on cluster masses for systems typically at $z\lesssim 1$ \citep{Tamura16,Ivezic19,MSE19,Cornwell22,Cornwell23}.}

For each FoF halo, IllustrisTNG  provides a list of subhalos from the Subfind algorithm \citep{subfind2001}.  For clusters in Table \ref{table:3dinfo}, we extract  subhalos with stellar mass  $ M_\star > 10^8$M$_\odot$ and within $10R_{200c}$ of the center of the cluster halo. We identify these subhalos as cluster  member galaxies.

\begin{table}[htbp]
\begin{center}
\caption{\label{table:3dinfo} Cluster samples from TNG300-1.}
\begin{tabular}{cccccc}
\hline
\hline
 $z$ & no of & median $M_{200c}$ & 50\% {perc.} & min-max\\
     & clusters & & range & $M_{200c}$\\
     &  & $\left[10^{14}~\text{M}_{\odot }\right]$  & $\left[10^{14}~\text{M}_{\odot }\right]$ & $\left[10^{14}~\text{M}_{\odot }\right]$ \\
\hline
 & & \\
0.01 & 282 & 1.59 & 1.22--2.33 & 1.00--15.0 \\
0.11 & 255 & 1.58 & 1.24--2.19 & 1.00--12.6\\
0.21 & 231 & 1.47 & 1.22--2.24 & 1.01--12.3 \\
0.31 & 201 & 1.50 & 1.20--2.21 & 1.01--13.6 \\
0.42 & 178 & 1.43 & 1.21--2.03 & 1.00--13.6 \\
0.52 & 145 & 1.41 & 1.20--2.01 & 1.01--12.1 \\
0.62 & 122 & 1.43 & 1.17--1.91 & 1.00--8.84 \\
0.73 & 98 & 1.35 & 1.15--1.90 & 1.00--8.92 \\
0.82 & 80 & 1.38 & 1.12--2.02 & 1.00--8.43 \\
0.92 & 60 & 1.38 & 1.19--1.92 & 1.00--7.60 \\
1.04 & 45 & 1.43 & 1.19--1.91 & 1.02--4.37 \\
\hline
 \end{tabular}
 \end{center}
 \end{table}

\begin{figure}
    \centering
    \includegraphics[width=\columnwidth]{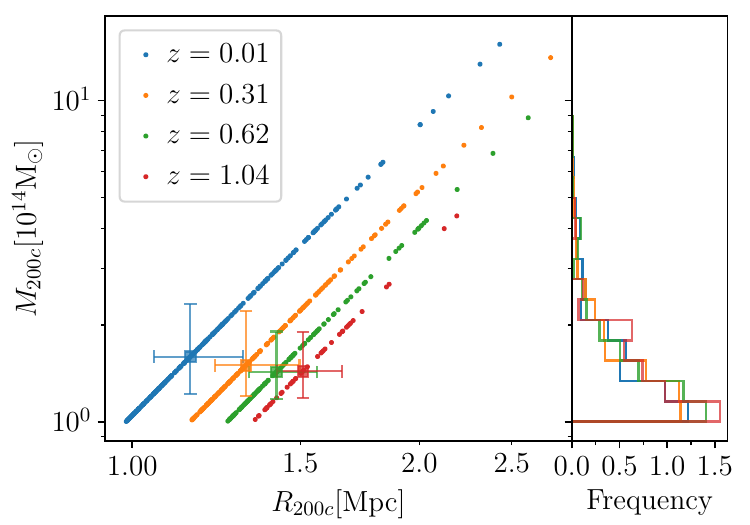}
    \caption{Relation between $M_{200c}$ and comoving $R_{200c}$ for IllustrisTNG clusters. Left panel: Blue, orange, green, and red points show the $M_{200c}-R_{200c}$ relation for clusters in four redshift bins $z=0.01,0.31,0.62$, and $z=1.04$, respectively. Colored squares with error bars show the median and interquartile range in each redshift bin. Right panel: Mass distribution histograms, with areas normalized to unity.}
    \label{fig:mr200}
\end{figure}

Blue, orange, green, and red points in Fig. \ref{fig:mr200} show the relation between $M_{200c}$ and the comoving $R_{200c}$ for clusters in four redshift bins: $z=0.01,0.31,0.62$, and $z=1.04$, respectively. Colored squares with error bars show the median and the interquartile range for each sample.

Figure \ref{fig:mr200} and Table \ref{table:3dinfo} show that, as the redshift increases from $z=0.01$ to $z=1.04$, the sample minimum $M_{200c}$ is nearly constant. The maximum $M_{200c}$ decreases by $71\%$ with increasing redshift. The lower and upper extrema of the interquartile ranges of $M_{200c}$ behave in a similar way. 

The mass functions in each bin are skewed towards low masses, $M_{200c}\lesssim 3\cdot 10^{14}$M$_\odot$, even at low redshift. Thus even at low redshift the high-mass tails have small statistical weight. As the redshift increases from $z=0.01$ to $z=1.04$, the median $M_{200c}$ of clusters decreases by only $10\%$. The comoving $R_{200c}$'s are proportional to $M_{200c}^{1/3}$ (Fig. \ref{fig:mr200}).

\subsection{Mass and velocity profiles}\label{subsec:profiles}

\citet{Pizzardo23} compute a cumulative mass profile for each cluster, $M(<r)$, based on the 3D distribution of matter extracted from raw snapshots. These profiles include all matter species: dark matter, gas, stars, and black holes. For each cluster, \citet{Pizzardo23} compute $M(<r)$ in 200 logarithmically spaced bins covering  the radial range $(0.1-10)R_{200c}^{3D}$. These profiles define $R_{200c}$ and $M_{200c}$ for each cluster; they allow straightforward computation of cumulative and shell density profiles for each cluster. 

\citet{Pizzardo23mar} compute a single average galaxy radial velocity profile at each redshift by averaging over all the individual galaxy radial velocity profiles for the clusters in the redshift bin. They use a subsample of the cluster sample we use here, because they include only the 78\% of systems allowing application of the caustic technique \citep{Diaferio1997,Diaferio99,Serra2011}, an observational method for estimating  the cluster mass profile. We include the entire set of IllustrisTNG clusters to maximize the sample size and  to minimize any potential biases in the mass function and resulting radial velocity profile.

We compute the galaxy radial velocity profile of individual clusters in each subsample. Based on the comoving position of simulated galaxies with respect to the cluster center, $\mathbf{r}_{c,i}$, and the galaxy peculiar velocity, $\mathbf{v}_{p,i}$, we compute the radial velocity of each galaxy: ${\rm v}_{{\rm rad},i}=[{\bf v}_{p,i} + H(z_s)a(z_s){\bf r}_{c,i}]\cdot {\bf r}_{c,i}/r_{c,i}$, where $H(z_s)$ and $a(z_s)$ are the Hubble function and the scale factor at the redshift $z_s$ of the snapshot. We compute the mean radial velocity profile of the cluster by  averaging over the galaxy ${\rm v}_{{\rm rad}}(r)$'s within 100 linearly spaced radial bins covering the range $(0,10)R_{200c}^{3D}$. At each redshift, we average over all of the individual radial velocity profiles for the snapshot. We obtain a single mean galaxy radial velocity profile along with the dispersion around it.

 \begin{figure*}
    \centering
    \includegraphics[width=\textwidth]{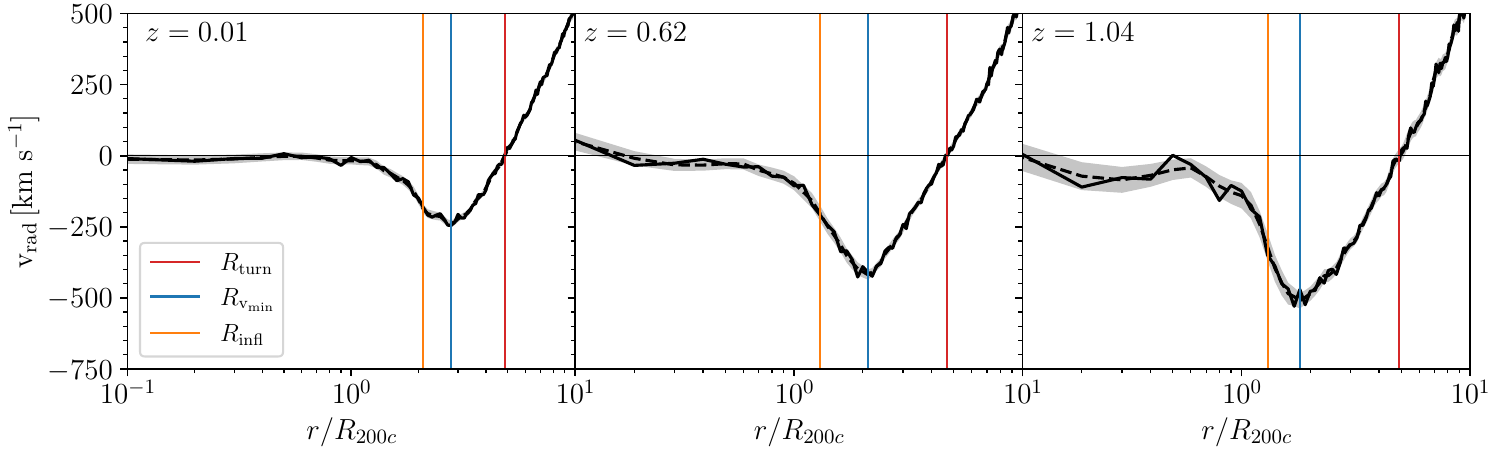}
    \caption{Average radial velocity profiles and dynamical radii. Solid curves show the average galaxy radial velocity profiles of clusters at three redshifts: $z=0.01,0.62$, and 1.04, from left to right, respectively. Dashed curves show the Savitzky-Golay \citep{Savitzky64} smoothed profiles. Shaded bands show the error in the smoothed profiles. In each panel, the red, blue, and orange lines show the turnaround, the minimum radial velocity, and the inflection point radii, respectively, derived from the average radial velocity profiles (see Sect. \ref{sec:radiigal}). }
    \label{fig:vprof}
\end{figure*}
The solid curves in Fig. \ref{fig:vprof} show the average galaxy radial velocity profiles for  clusters at  three redshifts $z=0.01,0.62$, and 1.04, from left to right, respectively. The dashed curves show  Savitzky-Golay \citep{Savitzky64} smoothed profiles derived with an {11 bin window and a fourth order polynomial interpolation}. Shaded bands show the error in the smoothed profiles. Fluctuations and errors increase with increasing redshift because of the decreasing size of the cluster samples (see second column of Table \ref{table:3dinfo}).

\section{Dynamical radii from radial velocity profiles}\label{sec:radiigal}

The galaxy radial velocity profile provides direct  measures of both the turnaround radius and the infall velocity minimum for a cluster of galaxies. The radial velocity profile also provides a complementary view of the splashback radius $R_{\rm spl}$ as the inner boundary of the region where infalling galaxies dominate the cluster dynamics. We derive  the turnaround radius and the point of minimum radial velocity in Sects. \ref{subsec:turn} and \ref{subsec:rvmin}, respectively. {Sect. \ref{subsec:rsplash}, shows how these dynamically determined radii provide an interpretation of $R_{\rm spl}$ based on the radial velocity profile. We compare the dynamically derived radii with $R_{\rm spl}^{\rm n_g}$, an observational proxy for $R_{\rm spl}$ } \citep[e.g.][]{adhikari2014,Diemer2014,More2015,Diemer17i,Diemer2017sparta2,Diemer18,oneil21}.

\subsection{The turnaround radius}\label{subsec:turn}

The turnaround radius $R_{\rm turn}$ of a galaxy cluster is {\rm defined as} the cluster-centric distance where galaxies depart from the Hubble flow \citep{Gunn1972,silk1974,Schechter80}. 
{Hence} $R_{\rm turn}$ is the radius where the {smoothed} galaxy radial velocity  ${\rm v_{rad}}(R_{\rm turn})=0$. The red vertical lines in Fig. \ref{fig:vprof} show the turnaround radius at redshifts $z=0.01,0.62$, and 1.04, respectively from left to right.

\begin{figure}[htbp]
    \centering
    \includegraphics[width=\columnwidth]{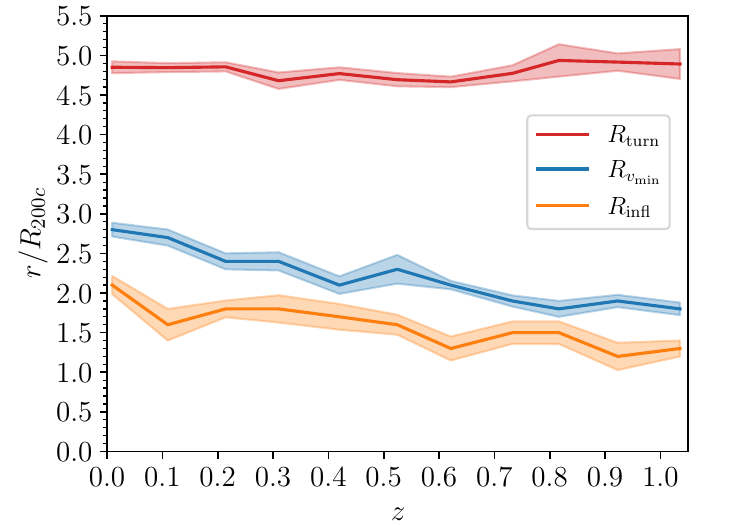}
    \caption{Dynamical radii from {smoothed} average galaxy radial velocity profiles. The red, blue, and orange lines show the mass averaged turnaround radius  (Sect. \ref{subsec:turn}), the point of minimum ${\rm v_{rad}}(r)$  (Sect. \ref{subsec:rvmin}), and the inflection point of ${\rm v_{rad}}(r)$ (Sect. \ref{subsec:rsplash}), as a function of redshift, respectively. Shadowed bands show the error from bootstrap resampling of the correspondingly colored radius.}
    \label{fig:radiigal_glob}
\end{figure}

The red line in Fig. \ref{fig:radiigal_glob} shows  $R_{\rm turn}$ in units of $R_{200c}$ as a function of redshift. The red shadowed band shows the uncertainty in $R_{\rm turn}$, based on bootstrapping 1000 samples  at each redshift. $R_{\rm turn}$ is in the range $(4.67-4.94)R_{200c}$. The turnaround radius is independent of redshift with a typical value $R_{\rm turn} = (4.81\pm 0.10)R_{200c}$. 

We compare the $R_{\rm turn}$ values derived from the average radial velocity profile with the analytic predictions of \citet[][private communication to \citet{villumsen86}]{Meiksin85}, $R_{\rm turn}^{\rm Meik}$.
{In this approach, the density contrast is $\delta(r) = 3M(r)/4\pi \rho_{bkg} r^3 - 1$, where $M(r)$ is the mass of the cluster within a distance $r$ from the cluster center and $\rho_{bkg}=\Omega_{m}\rho_c$ is the background matter density. In the spherical collapse model, the radial velocity induced by the perturbation is $v_{\rm rad}/Hr \approx \Omega_m^{0.6} P(\delta)$ \citep{Gunn1972,silk1974,Schechter80,Regos89}. Meiksin's approach approximates the function $P(\delta)$ with a non-polynomial; the overdensity within the turnaround radius where ${\rm v}_{\rm rad}/Hr=1$, is 
\begin{align}
    &\delta_{t, \rm Meik} = \frac{3}{2} \Omega_m^{-1.2}\left( 1+\sqrt{1+4\Omega_m^{1.2}} \right).
\end{align} }

At each redshift, we compute the set of $R_{\rm turn}^{\rm Meik}$'s from the mass profiles of individual clusters in Table \ref{table:3dinfo}. We adopt the \citet{Lahav1991} approximation for the growth factor. At each redshift, we compute the median value $\hat{R}_{\rm turn}^{\rm Meik}$ and the interquartile range.

$R_{\rm turn}$ derived from the average radial velocity profile is consistent with the analytic predictions of the Meiksin approximation. On average $R_{\rm turn}$ exceeds $\hat{R}_{\rm turn}^{\rm Meik}$ by $\lesssim 3.9\%$. At each redshift, $R_{\rm turn}$ is within the interquartile range of the set of individual $R_{\rm turn}^{\rm Meik}$'s, and $R_{\rm turn}$ and $\hat{R}_{\rm turn}^{\rm Meik}$ agree to within $\sim 1.5\sigma$. $\hat{R}_{\rm turn}^{\rm Meik}$ increases by $\sim 4\%$ as the redshift increases from $z=0.01$ to $z=1.04$; this increase is not ruled out by the IllustrisTNG results.

\subsection{The minimum radial velocity}\label{subsec:rvmin}

The point of minimum galaxy radial velocity, $R_{\rm v_{min}}$, is a  characteristic feature of the ${\rm v_{rad}}(r)$ profiles.  
Previous evaluations of  $R_{\rm v_{min}}$ include computations based on dark matter particles within stacked halo samples in N-body simulations \citep{Cuesta08,deBoni2016,pizzardo2020,Fong21,Gao23}, and derivations based on  the intracluster gas from hydrodynamical simulations \citep{Valles20}. {\citet{Fong21} and \citet{Gao23} identified $R_{\rm v_{min}}$ with the inner depletion radius, the radius where the mass flow is maximum.} \citet{Pizzardo23mar} also derive  $R_{\rm v_{min}}$ from galaxy velocities of clusters from the IllustrisTNG simulations for redshifts $0.01\leq z\leq 1.04$. 

The blue vertical lines in Fig. \ref{fig:vprof} show  $R_{\rm v_{min}}$ from the {smoothed} average galaxy radial velocity profiles at the three redshifts $z=0.01,0.62,$ and $1.04$, from left to right, respectively. The blue curve in Fig. \ref{fig:radiigal_glob} shows $R_{\rm v_{min}}$ as a function of redshift. $R_{\rm v_{min}}$  decreases with increasing redshift by $\sim 41\%$.
The redshift dependence occurs because at fixed mass, clusters at low redshift accrete from relatively less dense surroundings than clusters at higher redshift. The cores of high redshift clusters are also more dense relative to their surroundings, bringing the maximum infall velocity closer to the cluster center. The splashback radius that we consider next must lie within the radius where the radial velocity has its minimum.

\subsection{The splashback radius}\label{subsec:rsplash}

The splashback radius $R_{\rm spl}$, a proxy for the physical boundary of a galaxy cluster, is the average location of the first apocenter of matter recently accreted by the cluster halo. {Therefore $R_{\rm spl}$ marks the boundary where the internal cluster dynamics dominates systematic infall} \citep[e.g.][]{adhikari2014,Diemer2014,More2015,Diemer17i,Diemer2017sparta2,Diemer18,Diemer20,oneil21}. 

The observable proxy of $R_{\rm spl}$, $R_{\rm spl}^{\rm n_g}$, is the minimum of the logarithmic slope of the local matter density $\rho$, $\gamma=\frac{{\rm d}\log \rho(r)}{{\rm d}\log r}$ \citep{adhikari2014}. {We use simulated galaxies in IllustrisTNG to compute the average galaxy local number density ${\rm n_g}(r)$ at each redshift; we then derive the logarithmic slope of $n_{\rm n_g}(r)$, $\gamma_{\rm n_g}(r)$, and smooth it with a Savitzky-Golay filter \citep{Savitzky64} with an 11 bin window and a fourth order polynomial. The minimum of the resulting function is $R_{\rm spl}^{\rm n_g}$. }

{We check the stability of the results by considering filters with polynomials of order two to six and fixed 11 bin windows.  All of the corresponding $R_{\rm spl}^{\rm n_g}$s are within $\sim 5\%$ of $R_{\rm spl}^{\rm n_g}$s resulting from the fourth order polynomial; furthermore, $\chi^2$ yields very similar results for all filter choices.}

\begin{figure}[htbp]
    \centering
    \includegraphics[width=\columnwidth]{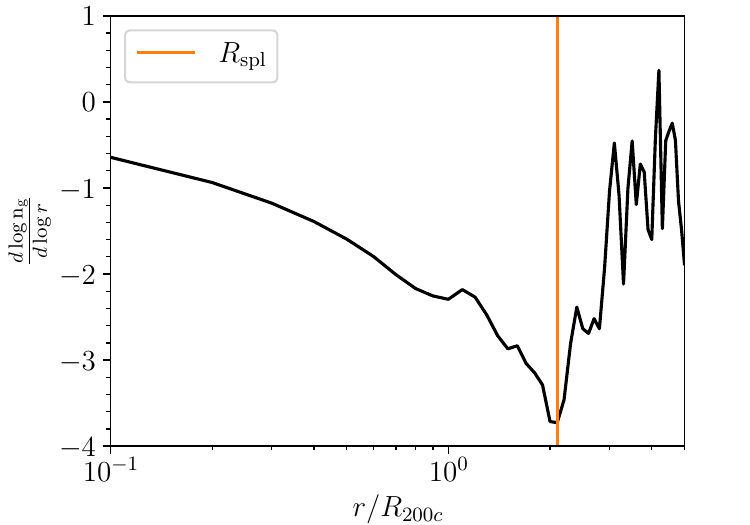}
    \caption{The splashback radius from ${\rm n_g}$. The curve shows the {smoothed logarithmic slope of the average local galaxy number density} of the 282 clusters at $z=0.01$. The orange vertical line shows the minimum, $R_{\rm spl}^{\rm n_g}$. }
    \label{fig:radiirho}
\end{figure}

Figure \ref{fig:radiirho} shows $\gamma_{\rm n_g}(r)$ based on the average galaxy local number density profile of the 282 TNG300-1 clusters with $M_{200c}>10^{14}$M$_\odot$ at $z=0.01$.  The orange vertical line shows the minimum of $\gamma_{\rm n_g}(r)$, $R_{\rm spl}^{\rm n_g}$, {an observationally accessible  proxy of $R_{\rm spl}$.}

\begin{figure}[htbp]
    \centering
    \includegraphics[width=\columnwidth]{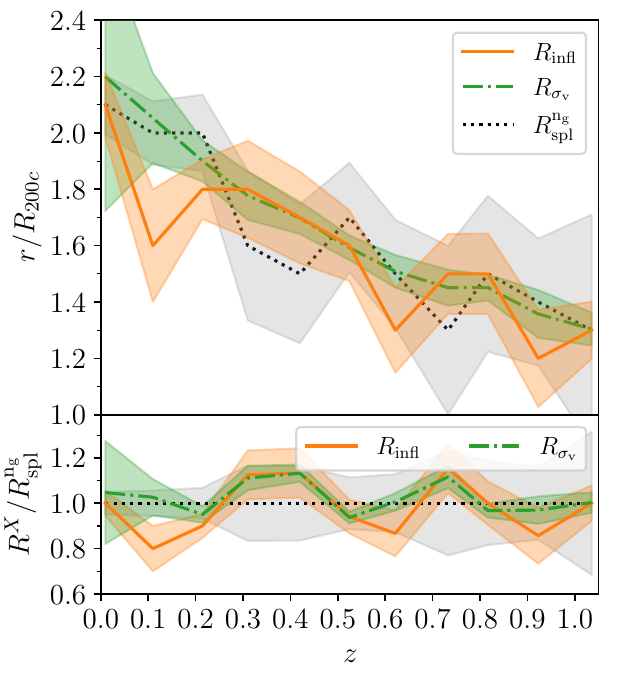}
    \caption{Dynamical radii from cluster galaxy velocity in the splashback region. Upper panel: The solid orange, dash-dotted green, and dotted black lines show $R_{\rm infl}$, $R_{\rm \sigma_v}$, and $R_{\rm spl}^{\rm n_g}$, respectively. The shadowed area shows the associated standard deviation. Bottom panel: The solid orange and dash-dotted green lines show the ratio between $R_{\rm infl}$ and $R_{\rm spl}^{\rm n_g}$, and between $R_{\rm \sigma_v}$ and $R_{\rm spl}^{\rm n_g}$, respectively. The dotted line shows equality. The orange, green, and grey shadowed areas show the 1$\sigma$ confidence levels for each ratio with the corresponding color.}
    \label{fig:splashback_defs}
\end{figure}

{The black dotted curve in the upper panel of Fig. \ref{fig:splashback_defs} shows the dependence of $R_{\rm spl}^{\rm n_g}$ on redshift. The gray shadowed area shows the standard deviation based on bootstrap  resampling. Over the redshift range we sample, $R_{\rm spl}^{\rm n_g}$ decreases  by $\sim 38\%$. At fixed mass $R_{\rm spl}$ should decrease with redshift (see Sect. \ref{sec:introduction}). The uncertainty increases with increasing redshift. At higher redshift, the size of the cluster sample is small (see Table \ref{table:3dinfo}); hence the $\gamma_{\rm n_g}(r)$ profile is very noisy. Locating the minima of these noisy $\gamma_{\rm n_g}(r)$'s is challenging and the uncertainty is correspondingly large. }

{The $R_{\rm spl}^{\rm n_g}$  we derive are consistent with recent results of \citet{oneil21} who also compute $R_{\rm spl}^{\rm n_g}$ with TNG300-1.  \citet{oneil21} express their  $R_{\rm spl}^{\rm n_g}$s in units of $R_{200m}$. We thus rescale our $R_{\rm spl}^{\rm n_g}$s by computing the $R_{200m}$ of the average matter density profiles. The ratio $R_{200m}/R_{200c}$ is in the range $1.63-1.65$ throughout the redshift range and is redshift independent  We thus rescale  $R_{\rm spl}^{\rm n_g}$s with the average ratio $R_{200m}/R_{200c}=1.64$. }

{The $R_{\rm spl}^{\rm n_g}$s we derive agree with those of \citet{oneil21} at $z\leq 0.62$; they are $\sim 20-50\%$ larger at higher redshifts. However, the samples we extract are not fully comparable with theirs: we define galaxies as subhalos with stellar mass $M_\star > 10^8$M$_\odot$ (see Sect. \ref{subsec:cats}); they define galaxies by selecting  a total subhalo mass $>10^9$M$_\odot$ and nonzero $M_\star$. Thus, the physical properties of the two samples may differ; the choice we make reduces the number of subhalos labelled as galaxies leading to poorer statistics.\footnote{\citet{oneil21} compute $R_{\rm spl}^{\rm n_g}$ by fitting the density profiles with an analytic function. They show that different fitting procedures can affect the splashback radius by $\lesssim 8\%$. The agreement between the $R_{\rm spl}^{\rm n_g}$s we derive and those of \citet{oneil21} at $z\lesssim 0.62$ suggests that our computation of $R_{\rm spl}^{\rm n_g}$ is robust.  Mild discrepancies at high redshift result from different sampling of the simulated galaxy population.}. The redshift dependence of $R_{\rm spl}^{\rm n_g}$s we obtain is consistent with previous results based on N-body simulations \citep{More2015,Diemer20,oneil21}, although with a higher normalization.}

The average galaxy radial velocity profile based on IllustrisTNG provides a route to define two new radii which closely approximate $R_{\rm spl}^{\rm n_g}$, thus providing additional physical insight into galaxy dynamics at $R_{\rm spl}$. We first explore a radius based on the inflection point in the radial velocity profile, $R_{\rm infl}$. We then turn to a radius based on a comparison between the velocity dispersion and the mean radial velocity, $R_{\rm \sigma_v}$. In both cases these radii mark the limiting cluster-centric radius where infall no longer dominates the cluster dynamics.

The radial velocity profiles (Fig. \ref{fig:vprof}) turn from concave $\left(\frac{{\rm d}^2\mathrm{v_{rad}}}{{\rm d}r^2}<0\right)$ to convex $\left(\frac{{\rm d}^2\mathrm{v_{rad}}}{{\rm d}r^2}>0\right)$ as the cluster-centric radius increases. The inflection point where the change occurs, $R_{\rm infl}$, lies between $R_{200c}$ and $R_{\rm v_{min}}$. 
The radius of the inflection point is in the range $\sim (2-3)R_{200c}$ depending on redshift (see Sect. \ref{subsec:rvmin}) and is evident at every redshift. Statistically, galaxies within the inflection point cannot escape to larger radii. Thus the inflection point of ${\rm v_{rad}}(r)$ is a dynamically motivated radius {that should approximate $R_{\rm spl}$.}

Computationally, the radius corresponding to the inflection is the  minimum of $\frac{{\rm d}\mathrm{v_{rad}}}{{\rm d}r}$. Here ${\rm v_{rad}}(r)<0$. The radial velocity also increases in absolute value as the cluster-centric distance increases. Thus, at the inflection point, the local spatial change of ${\rm v_{rad}}(r)$ is a maximum in absolute value.

\begin{figure}
    \centering
    \includegraphics[width=\columnwidth]{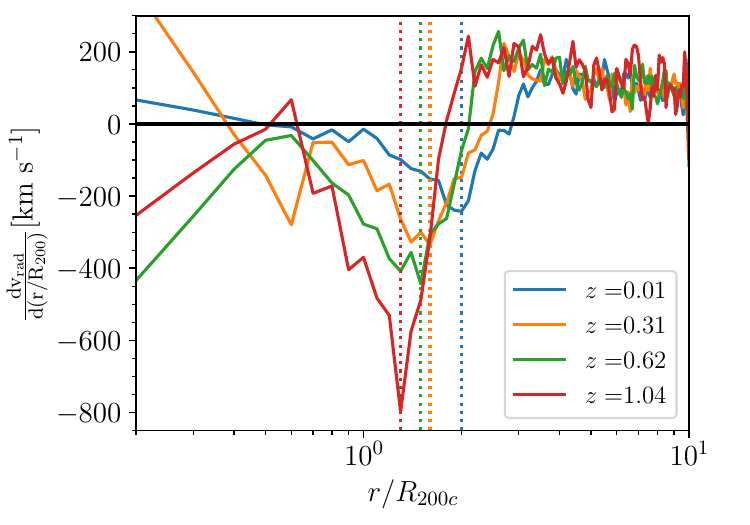}
    \caption{Radial derivative of {the smoothed} $\mathrm{v_{rad}}(r)$, and $R_{\rm infl}$. The blue, orange, green, and red curves show the smoothed $\frac{{\rm d}\mathrm{v_{rad}}}{{\rm d}r}$ profiles for the four redshifts $z=0.01,0.31,0.62$, and 1.04, respectively. The vertical lines show the minimum $R_{\rm infl}$ correspondingly to each colored curve.}
    \label{fig:dvrad}
\end{figure}

To measure $R_{\rm infl}$, we first compute $\frac{{\rm d}\mathrm{v_{rad}}}{{\rm d}r}$ at each redshift from the smoothed average galaxy radial velocity profile. {The colored curves in Fig. \ref{fig:dvrad} show the resulting $\frac{{\rm d}\mathrm{v_{rad}}}{{\rm d}r}$ profiles for the four redshifts listed in the legend.} At each redshift, the minimum of the corresponding profile locates $R_{\rm infl}$. The dotted vertical lines in Fig. \ref{fig:dvrad} indicate the minimum $R_{\rm infl}$ for the correspondingly coloured curves.

For comparison with other dynamical radii, the orange vertical lines in Fig. \ref{fig:vprof} show the inflection point $R_{\rm infl}$ at the three redshifts $z=0.01,0.62,$ and 1.04, from left to right, respectively. The orange curve in Fig. \ref{fig:radiigal_glob} summarizes the behavior of  $R_{\rm infl}$ as a function of redshift. The shadowed area indicates the standard deviation based on bootstrap resampling. From $z=0.01$ to $z=1.04$, $R_{\rm infl}$ decreases from $\sim 2.1 R_{200c}$ to $\sim 1.3 R_{200c}$ ($\sim 38\%$). On average, $R_{\rm infl}$ is $\sim 28\%$ smaller than $R_{\rm v_{min}}$.

\begin{figure}
    \centering
    \includegraphics[width=\columnwidth]{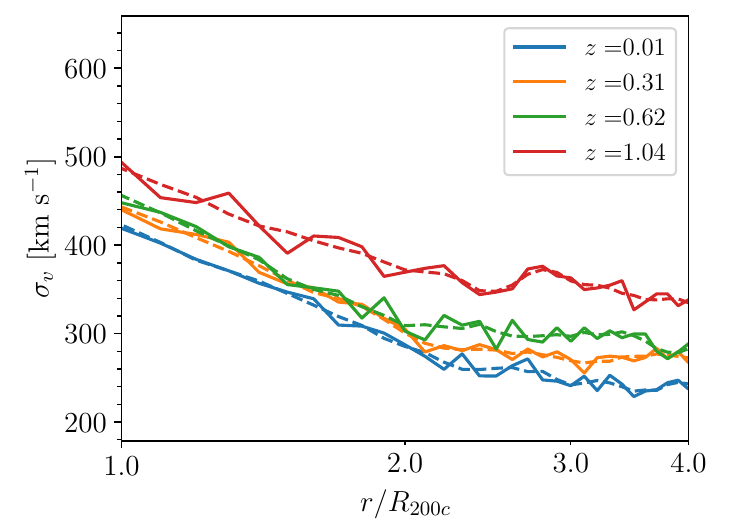}
    \caption{Average galaxy velocity dispersion as a function of cluster-centric radius. The blue, orange, green, and red solid curves show the dispersion at four redshifts $z=0.01, 0.31, 0.62$, and 1.04, respectively. {The dashed curves show the smoothing of the correspondingly colored solid curves.} }
    \label{fig:sigmav}
\end{figure}

A second radius that improves understanding of the physics within the clusters' splashback region is $R_{\rm \sigma_v}$, which we define from the average ratio between the 3D cluster velocity dispersion and the infall velocity, $\sigma_{\rm v}(r)/{\rm v_{rad}}(r)$. 
{Figure \ref{fig:vprof} shows that from $R_{\rm turn}$ to $R_{\rm v_{min}}$, ${\rm v_{rad}}$ increases in absolute value from $\sim 0$ to $\sim 250-500$~km~s$^{-1}$ depending on the redshift. Then, from $R_{\rm v_{min}}$ to $R_{200c}$, ${\rm v_{rad}}$ decreases in absolute value from its maximum to $\sim 0$. Fig. \ref{fig:sigmav} shows that $\sigma_{\rm v}(r)$  increases monotonically from $R_{\rm turn}$ to $R_{200c}$; $\sigma_{\rm v}(r)$ increases from $\sim 250-350$~km~s$^{-1}$ to $\sim 400-500$~km~s$^{-1}$. At $R_{\rm v_{min}}$, the ratio between $\sigma_{\rm v}(r)$ and ${\rm v_{rad}}$ is always greater than $-1$: from $z=0.01$ to $z=0.52$ the ratio increases from $\sim-0.97$ to $\sim 0.74$; at higher redshifts, the ratio at $R_{\rm v_{min}}$ is $\sim -0.72$. The two profiles $\sigma_{\rm v}(r)$ and ${\rm v_{rad}}$ always have two intersections. When $\sigma_{\rm v}(r)>|{\rm v_{rad}}(r)|$, randomly oriented motions exceed  motions aligned along the radial direction; radial motions dominate  when $\sigma_{\rm v}(r)<|{\rm v_{rad}}(r)|$.} 
We compute $R_{\rm \sigma_v}$ by identifying the smallest cluster-centric radius where $ \sigma_{\rm v}(r)/{\rm v_{rad}}(r)=-1$. Within this radius galaxies orbiting in the cluster potential dominate the dynamics.

At each redshift, we compute the average galaxy velocity dispersion profile $\sigma_{\rm v}(r)$ with an approach similar to the computation of the average ${\rm v_{rad}}(r)$ profiles (Sect. \ref{subsec:profiles}). We consider 100 shells with linearly spaced boundaries covering the range $(0-10)R_{200c}$. For each cluster, we compute the standard deviation of the velocities of all the galaxies inside that shell, the estimator of the velocity dispersion for the shell. In  bin $n$ for the shell with radial range $(r_n-r_{n+1})$, the velocity dispersion is the square root of
\begin{equation}
    {\rm \sigma_v}^{(n)^{\,{2}}} = \sum_{\substack{\text{galaxies $i=1,...,N$} \\ \text{with } r_n< r/R_{200c} \leq r_{n+1}}} \;  \sum_{k=x,y,z}  \frac{\left(\mathrm{v}_k^{(i)}-\bar{\rm v}_k^{(n)}\right)^2}{3N},  
\end{equation}
where $k$ runs over the three spatial components, $\bar{\rm v}_k^{(n)}$ is the mean of the $k$ spatial component of the velocity of galaxies in bin $n$, and $N$ is the number of galaxies.
The single average galaxy $\sigma_{\rm v}(r)$ profile at each redshift is the mean of the velocity dispersions for the  individual clusters in each bin. 

{The curves in Fig. \ref{fig:sigmav} show the average $\sigma_{\rm v}(r)$ for clusters at the four redshifts listed in the legend} within the radial range $(1-4)R_{200c}$. Generally, $\sigma_{\rm v}(r)$ deceases with increasing cluster-centric radius. The $\sigma_{\rm v}(r)$ profile at fixed cluster-centric radius increases with redshift by  $\sim 40\%$  from $z=0.01$ to $z=1.04$.

\begin{figure}[htbp]
    \centering
    \includegraphics[width=\columnwidth]{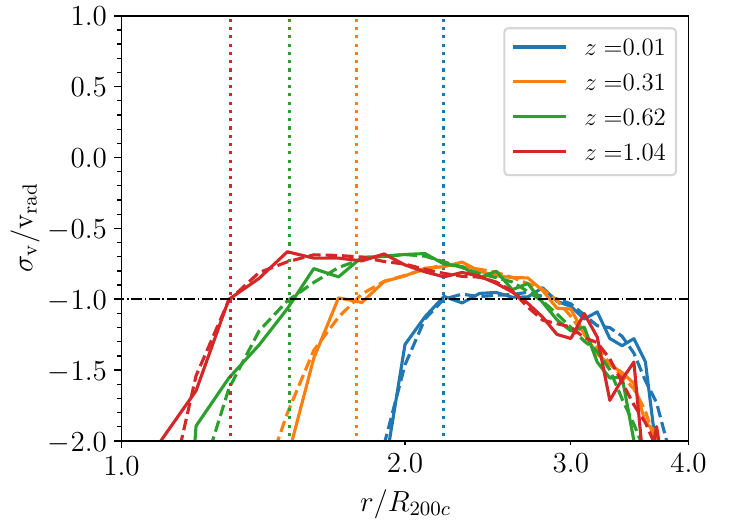}
    \caption{Average ratio between the galaxy velocity dispersion and the galaxy radial velocity, as a function of cluster-centric radius. The blue, orange, green, and red solid curves show the ratio at four redshifts $z=0.01, 0.31, 0.62$, and 1.04, respectively. {The dashed curves show the smoothed correspondingly colored solid curves.} The horizontal dash-dotted line shows $\sigma_{\rm v}/{\rm v_{rad}}= -1$. The vertical dotted lines show $R_{\rm \sigma_v}$ corresponding to $\sigma_{\rm v}(r)/{\rm v_{rad}}(r)= -1$ {from smoothed profiles}.}
    \label{fig:ratio_sigmavrad}
\end{figure}

Figure \ref{fig:ratio_sigmavrad} shows the ratio $ \sigma_{\rm v}(r)/{\rm v_{rad}}(r)$ for {the four redshifts listed in the legend}. As in Fig. \ref{fig:sigmav}, we limit the radial range to $(1-4)R_{200c}$.

At cluster-centric distances $\lesssim R_{200c}$, where the cluster is in approximate dynamical equilibrium, ${\rm \sigma_v}(r)/{\rm v_{rad}}(r)$ is ill-defined because ${\rm v_{rad}}(r) \sim 0$ (Fig. \ref{fig:vprof}). At radii $\sim (1-2)R_{200c}$,  ${\rm \sigma_v}(r)/{\rm v_{rad}}(r)$ increases but remains $ < -1$. This behavior occurs at smaller radii with increasing  redshift. In this radial range there is a net radial motion toward the cluster center, but the infall does not dominate over the internal cluster dynamics. 

In the region where $\sigma_{\rm v}(r)/{\rm v_{rad}}(r)>-1$ (Fig. \ref{fig:ratio_sigmavrad}), radial infall dominates the dynamics. The center of this range is $\approx R_{\rm v_{min}}$ as expected (Sect. \ref{subsec:rvmin}). Statistically, galaxies in this region have radial velocities directed toward the cluster center (Fig. \ref{fig:vprof}); the radial velocities exceed the orbital velocities. At smaller cluster-centric radii, where $\sigma_{\rm v}(r)/{\rm v_{rad}}(r)<-1$, orbital motions dominate. Once galaxies migrate from the infall region where $\sigma_{\rm v}(r)/{\rm v_{rad}}(r)>-1$ into the inner cluster region where $\sigma_{\rm v}(r)/{\rm v_{rad}}(r)<-1$, they can generally no longer reach radii where  $\sigma_{\rm v}(r)/{\rm v_{rad}}(r)> -1$. Thus the radius where $\sigma_{\rm v}(r)/{\rm v_{rad}}(r)=-1$, $R_{\rm \sigma_v}$, {provides another complementary physical view of the dynamics near the splashback radius.}

{$R_{\rm \sigma_v}$ is the innermost cluster-centric radius where $\sigma_{\rm v}(r)/{\rm v_{rad}}(r)=-1$. To locate $R_{\rm \sigma_v}$, we use the ratio between the smoothed $\sigma_{\rm v}(r)$ and ${\rm v_{rad}}(r)$ profiles.} The dotted vertical lines in Fig. \ref{fig:ratio_sigmavrad} show $R_{\rm \sigma_v}$ for the correspondingly colored ratios. The green dash-dotted curve in the upper panel of Fig. \ref{fig:splashback_defs} shows that $R_{\rm \sigma_v}$ decreases with redshift. The green shadowed area indicates the error based on bootstrap resampling. From $z=0.01$ to $z=1.04$, $R_{\rm \sigma_{\rm v}}$ decreases from $\sim 2.2 R_{200c}$ to $\sim 1.3 R_{200c}$ ($\sim 40\%$).
 
At radii $\gtrsim 3R_{200c}$, ${\rm \sigma_v}(r)/{\rm v_{rad}}(r)$ decreases. Although ${\rm \sigma_v}(r)$ is roughly constant, the magnitude of ${\rm v_{rad}}(r)$ decreases as galaxies approach $R_{\rm turn}$ where they are just overcoming the Hubble flow and their net  radial velocity is zero.  Near the turnaround radius ($\sim (4.5-5)R_{200c}$, Sect. \ref{subsec:turn}), ${\rm v_{rad}}(r)\sim 0$; $\sigma_{\rm v}(r)/{\rm v_{rad}}(r)$ then behaves erratically.

Figure \ref{fig:splashback_defs} compares the observable proxy for the splashback radius, $R_{\rm spl}^{\rm n_g}$, with the two radii $R_{\rm infl}$ and $R_{\rm \sigma_v}$. These radii, derived from the mean radial velocity profile, mark the transition from the infall region  to the approximately virialized region where orbital motions dominate. 

In the upper panel of Fig. \ref{fig:splashback_defs}, the solid orange, dash-dotted green, and dotted black lines show $R_{\rm infl}$ (same as the orange line in Fig. \ref{fig:radiigal_glob}), $R_{\rm \sigma_v}$, and $R_{\rm spl}^{\rm n_g}$, respectively, as a function of redshift. The shadowed bands show the bootstrapped error. In the bottom panel, the solid orange and dash-dotted green lines show the ratios  $R_{\rm infl}/R_{\rm spl}^{\rm n_g}$ and $R_{\rm \sigma_v}/R_{\rm spl}^{\rm n_g}$, respectively. The orange, green, and grey shadowed areas show the 1$\sigma$ confidence levels for  $R_{\rm infl}$, $R_{\rm \sigma_v}$, and $R_{\rm spl}^{\rm n_g}$, respectively.

{Averaged over redshift, $R_{\rm infl}/R_{\rm spl}^{\rm n_g}= 1.00\pm 0.12$; $R_{\rm \sigma_v}/R_{\rm spl}^{\rm n_g}= 1.006\pm 0.069$ (bottom panel, Fig. \ref{fig:splashback_defs}). The redshift dependence of $R_{\rm spl}^{\rm n_g}$ agrees with that of $R_{\rm infl}$ and $R_{\rm \sigma_v}$ over the whole redshift range.}

The radii from the velocity profile, $R_{\rm infl}$ and $R_{\rm \sigma_v}$, have nearly the same value as $R_{\rm spl}^{\rm n_g}$. {The dynamically derived radii are unbiased and within $\sim 1\sigma$ of $R_{\rm spl}^{\rm n_g}$. Although $R_{\rm infl}$ and $R_{\rm \sigma_v}$ are not directly observable, they establish a connection between the splashback physics and the infall dynamics of clusters. The near equality of $R_{\rm infl}$, $R_{\rm \sigma_v}$ and $R_{\rm spl}^{\rm n_g}$ suggests  that $R_{\rm spl}$ corresponds to the boundary between the region where galaxies orbiting in the cluster potential dominate the dynamics and the region where  infall of galaxies dominates.}

\section{Discussion}\label{ref:discussion}

\subsection{The Cluster Mass Distribution}\label{subsec:massfz}

Table \ref{table:3dinfo} shows that the median cluster mass generally decreases as the redshift increases because very massive systems are progressively less abundant at higher redshift. The changing distribution of cluster masses could affect the resulting dynamical radii. We explore this issue by selecting subsamples that are homogeneous.

For each redshift, we construct homogeneous samples that include clusters with mass $ > M_{200c}$ in the range $(1.0-5.0)\cdot 10^{14}$~M$_\sun$. The last column of Table \ref{table:3dinfo} shows that this mass range is sampled at every redshift. A  Kolmogorov-Smirnov test demonstrates that the clipped samples have indistinguishable mass distributions; the p-values are in the range $(0.32-1.00)$. For each redshift, we compute the average galaxy radial velocity and the average galaxy velocity dispersion profiles for the clipped sample following  Sects. \ref{subsec:profiles} and \ref{subsec:rsplash}, respectively. 

The clipped $R_{\rm turn}$ are on average $\sim 0.1\%$ smaller than the full sample $R_{\rm turn}$, and they are always within $\lesssim 0.5\%$ of the full sample values. The difference between clipped and full sample $R_{\rm turn}$ as a function of redshift is unbiased. The clipped $R_{\rm v_{min}}$'s are equal to the full samples $R_{\rm v_{min}}$'s at every redshift except  $z=0.73$ where the clipped radius  exceeds the full sample radius by $\sim 5\%$. 

{The clipped $R_{\rm infl}$'s are equal to the full sample $R_{\rm infl}$'s at every redshift except $z=0.62$ where the clipped radius exceeds the full sample radius by $\sim 8\%$. The clipped $R_{\rm \sigma_v}$'s are on average $\sim 0.2\%$ below the full sample $R_{\rm \sigma_v}$'s and they lie within $\lesssim 0.8\%$ of the full sample $R_{\rm \sigma_v}$'s. Differences between the clipped and full sample $R_{\rm \sigma_v}$'s are unbiased. }
The dynamical radii are all insensitive to differences among the distribution of cluster masses at different redshifts in the full samples.

\subsection{Comparison of Galaxy and Total Matter Distribution Radii }\label{subsec:bias}

Galaxies may be biased tracers of the underlying distribution of  matter in the Universe, mainly comprised of dark matter \citep[e.g.,][]{Kaiser84,Davis85,White87}. IllustrisTNG enables
measurement of  the possible bias in the values of the dynamical radii derived  from the radial velocity profiles based on galaxies by comparing them with the values derived for the total matter distribution.

We compute a single average radial velocity for the total matter distribution at each redshift with the approach of Sect. \ref{subsec:profiles}. The total matter content includes dark matter, gas, stars, and black holes, which are all included in IllustrisTNG (see Sect. \ref{subsec:cats}). We take 200 logarithmically spaced bins  (rather than the 100 we used for the velocity profile based on galaxies alone) in the radial range $(0.1-10)R_{200c}$. The total matter distribution enables the more finely spaced bins.

\begin{figure}
    \centering
    \includegraphics[width=\columnwidth]{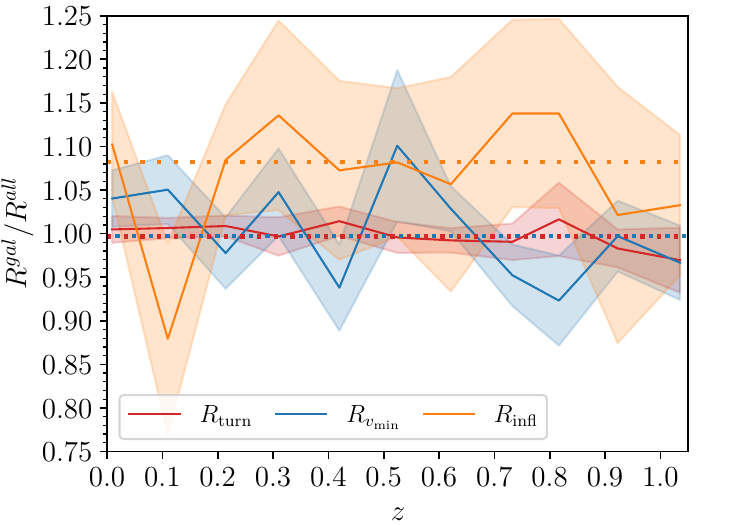}
    \caption{Ratio between dynamical radii from radial velocity profiles based on  galaxies and on the total matter. The red, blue, and orange lines show the ratio between the turnaround, the minimum radial velocity, and the inflection point of average radial velocity profiles of galaxies and total matter, as a function of redshift. The horizontal dotted lines show the median ratio of the corresponding curve for the 11 redshift bins. The shadowed band in each case  shows the $1\sigma$ confidence range.}
    \label{fig:gal_all_diff}
\end{figure}

At each redshift, we locate the turnaround, the minimum radial velocity, and the inflection point of the average total matter radial velocity profiles (Sect. \ref{sec:radiigal}). The radii derived from the total mass distribution  are $R_{\rm turn}^{\rm all}$, $R_{\rm v_{min}}^{\rm all}$, and $R_{\rm infl}^{\rm all}$, respectively.
The red, blue, and orange lines in Fig. \ref{fig:gal_all_diff} show the ratios between the galaxy based and total matter based radii:  $R_{\rm turn}/R_{\rm turn}^{\rm all}$, $R_{\rm v_{min}}/R_{\rm v_{min}}^{\rm all}$, and $R_{\rm infl}/R_{\rm infl}^{\rm all}$ as a function of redshift. The horizontal dotted lines show the corresponding median ratio for the 11 redshift bins. The  shadowed band of  corresponding color  shows  the $1\sigma$ confidence range for each  of the ratios.

{The radii $R_{\rm turn}$ and $R_{\rm v_{min}}$ based on galaxy velocities are consistent with the radii based on the total matter velocities.} Averaging over redshift, the median $R_{\rm turn}/R_{\rm turn}^{\rm all}\approx 0.997$ (red dotted horizontal line); $R_{\rm turn}$ is always within $\lesssim 4\%$ of $R_{\rm turn}^{\rm all}$. On average, $R_{\rm v_{min}}/R_{\rm v_{min}}^{\rm all}\approx 0.997$  (blue dotted horizontal line), and $R_{\rm v_{min}}$ overestimates (underestimates) $R_{\rm v_{min}}^{\rm all}$ by  $\lesssim +10\%$ ($-8\%$) at most. 
{The $R_{\rm infl}$'s based on galaxy velocities generally  exceed the corresponding radii based on the total matter. Averaging over redshift, $R_{\rm infl}/R_{\rm infl}^{\rm all}\approx 1.082$ (orange dotted horizontal line), and $R_{\rm infl}$ overestimates (underestimates) $R_{\rm infl}^{\rm all}$ by $\lesssim +14\%$ ($-12\%$). The results are consistent with equality at $\sim 1\sigma$ level.}
Figure \ref{fig:gal_all_diff} shows that the ratios are essentially independent of redshift. 

The mild overestimation of $R_{\rm infl}$ relative to $R_{\rm infl}^{\rm all}$ is consistent with the behaviour of $R_{\rm spl}$ found by \citet{oneil21} who also analyze TNG300-1. They compute the proxy for $R_{\rm spl}$ based on galaxies, $R_{\rm spl}^{\rm n_g}$, and compare it with the analogous splashback radius based on all of the matter, $R_{\rm spl}^{\rm n_{all}}$. {For systems with masses $\gtrsim 10^{14}$M$_\odot$, $R_{\rm spl}^{\rm n_g}/R_{\rm spl}^{\rm n_{all}} \sim 1.02-1.03$, with a considerable dispersion that allows overestimations as large as $\sim 15\%$ (see their Fig. 6). For the inflection point, the average ratio over the 11 redshift bins we sample is $R_{\rm infl}/R_{\rm infl}^{\rm all}=1.082 \pm 0.074$. This value is consistent with \citet{oneil21} within the uncertainty.} We conclude that, as \citet{oneil21} show for the splashback proxy, the bias between dynamical radii based on galaxies and all of the matter is small.

\section{Conclusion}\label{sec:conclusion}

Simulated galaxies  drawn from the IllustrisTNG300-1 simulations \citep{Pillepich18,Springel18,Nelson19} enable exploration of the infall region of 1697 galaxy clusters  with $M_{200c}>10^{14}$M$_\odot$ and redshift $0.01\leq z \leq 1.04$ \citep{Pizzardo23,Pizzardo23mar}. For these systems, we revisit the classical turnaround radius $R_{\rm turn}$ that defines the outer boundary of a cluster \citep{Gunn1972,silk1974,Schechter80} and the characteristic minimum infall velocity $R_{\rm v_{min}}$ \citep{deBoni2016,Valles20,Fong21,pizzardo2020,Pizzardo23mar}. Based on the galaxy radial velocity profile, ${\rm v_{rad}}(r)$, we derive two new measures of the inner boundary of the infall region. Both of these radii lie within the radial velocity minimum and coincide with the splashback radius, $R_{\rm spl}$ \citep{adhikari2014,Diemer2014,More2015}.

The  galaxy average radial velocity profile, ${\rm v_{rad}}(r)$, enables identification of the turnaround radius, $R_{\rm turn}$, where galaxies decouple from the Hubble flow.  The $R_{\rm turn}$'s  lie in the range $(4.67-4.94)R_{200c}$, are insensitive to redshift, and agree with the Meiksin analytic approximation \citep{Meiksin85}. 

The galaxy average radial velocity profile, ${\rm v_{rad}}(r)$, has a well-defined minimum, $R_{\rm v_{min}}$. The value of $R_{\rm v_{min}}$ decreases from $2.8 R_{200c}$ at $z=0.01$ to $1.8 R_{200c}$ at $z=1.04$. The maximum infall velocity itself increases with redshift.

Inside $R_{\rm v_{min}}$, we develop two new  dynamical radii that mark the inner boundary  of the infall region: (i) $R_{\rm infl}$ is the inflection point of ${\rm v_{rad}}(r)$, the cluster-centric radius where the derivative of ${\rm v_{rad}}(r)$ is maximum in absolute value, and (ii) $R_{\rm \sigma_v}$ is the smallest cluster-centric radius where $\sigma_{\rm v}(r) = |{\rm v_{rad}}(r)|$. Outside these radii, infall dominates the cluster dynamics; within these radii orbital motions dominate over infall.

To within 1$\sigma$, the  two dynamical radii, $R_{\rm infl}$  and  $R_{\rm \sigma_v}$, coincide with $R_{\rm spl}^{\rm n_g}$, the radius where the derivative of the galaxy average number density profile has a minimum. This radius is an observable proxy for the splashback radius, $R_{\rm spl}$, often determined from N-body simulations \citep{Diemer2014,Diemer2017sparta2,Mansfield17,Diemer18,Xhakaj2020}. Averaged over redshift, $R_{\rm infl}/R_{\rm spl}^{\rm n_g} = 1.00\pm 0.12$, and $R_{\rm \sigma_v}/R_{\rm spl}^{\rm n_g}= 1.006\pm 0.069$.
 
Galaxies may be biased tracers of the total matter content of  a cluster. For the set of IllustrisTNG clusters, the ratios between galaxy and total matter $R_{\rm turn}$ is $0.997\pm 0.014$; the corresponding ratio for $R_{\rm v_{min}}$ is $0.997\pm 0.056$.
$R_{\rm infl}$ for galaxies exceeds the value for all matter by $(8.2\pm 7.4)\%$. \citet{oneil21} measure a similarly negligible bias between values of $R_{\rm spl}$ for galaxies and the entire matter distribution.

The consistency between the dynamical radii, $R_{\rm infl}$  and  $R_{\rm \sigma_v}$,  and the splashback radius, $R_{\rm spl}^{\rm n_g}$,  provides an enhanced physical view of $R_{\rm spl}^{\rm n_g}$ as the inner boundary of the infall region {where galaxies are infalling onto the cluster core}.  Currently, $R_{\rm spl}^{\rm n_g}$ provides a direct observational route to this physical cluster boundary. {In the future, observations that include accurate independent distance measures  for galaxies around clusters and their filamentary structures will allow observational determination of the velocity field  for infalling galaxies \citep{Odekon22}. Some limited observations are available in the nearby Universe; they should become more broadly available at greater and greater depths in the Universe. Thus observational proxies for the dynamical radii $R_{\rm infl}$  and  $R_{\rm \sigma_v}$ should eventually complement determination of $R_{\rm spl}$ from the cluster density profile. }

\begin{acknowledgements} 
{We thank the referee for comments that led to improvements in this paper.} We thank Jubee Sohn for insightful discussions. 
M.P. and I.D. acknowledge the support of the Canada Research Chair Program and the Natural Sciences and Engineering Research Council of Canada (NSERC, funding reference number RGPIN-2018-05425).
The Smithsonian Institution supports the research of M.J.G. and S.J.K.
Part of the analysis was performed with the computer resources of INFN in Torino and of the University of Torino.
This research has made use of NASA's Astrophysics Data System Bibliographic Services.\\
All of the primary TNG simulations have been run on the Cray XC40 Hazel Hen supercomputer at the High Performance Computing Center Stuttgart (HLRS) in Germany. They have been made possible by the Gauss Centre for Supercomputing (GCS) large-scale project proposals GCS-ILLU and GCS-DWAR. GCS is the alliance of the three national supercomputing centres HLRS (Universitaet Stuttgart), JSC (Forschungszentrum Julich), and LRZ (Bayerische Akademie der Wissenschaften), funded by the German Federal Ministry of Education and Research (BMBF) and the German State Ministries for Research of Baden-Wuerttemberg (MWK), Bayern (StMWFK) and Nordrhein-Westfalen (MIWF). Further simulations were run on the Hydra and Draco supercomputers at the Max Planck Computing and Data Facility (MPCDF, formerly known as RZG) in Garching near Munich, in addition to the Magny system at HITS in Heidelberg. Additional computations were carried out on the Odyssey2 system supported by the FAS Division of Science, Research Computing Group at Harvard University, and the Stampede supercomputer at the Texas Advanced Computing Center through the XSEDE project AST140063.
\end{acknowledgements}

\bibliographystyle{aa}
\bibliography{main}

\end{document}